\documentclass[
 reprint,
 amsmath,
 amssymb,
 prd,
]{revtex4-2}

\usepackage{latexsym}
\usepackage[usenames]{color}
\usepackage{fancybox}
\usepackage{simplewick}
\usepackage{comment}
\usepackage{framed}
\definecolor{shadecolor}{rgb}{0.9,0.9,0.95}
\usepackage{setspace}
\usepackage{tabularx,booktabs}
\usepackage{bm}
\definecolor{darkgreen}{rgb}{0,0.5,0}
\definecolor{darkblue}{cmyk}{0.9,0.9,0,0}
\definecolor{darkred}{rgb}{0.6,0,0.3}

\usepackage{graphicx}
\usepackage[setpagesize=false,pagebackref=false, linktocpage, bookmarksopen=true, colorlinks=true, linkcolor=darkblue,citecolor=darkblue,urlcolor=darkblue]{hyperref}
\usepackage{hyperref}

\def\eqref#1{(\ref{#1})}

\def\beq{\begin{equation}}
\def\eeq{\end{equation}}

\begin{document}
\title{The four-point function of determinant operators in $\mathcal{N}=4$ SYM}

\author{Edoardo Vescovi}

\affiliation{
\vspace{5mm}
Nordita
\\
KTH Royal Institute of Technology and Stockholm University
\\
Roslagstullsbacken 23, SE-106 91 Stockholm, Sweden
}

\begin{abstract}
We calculate the four-point function of $1/2$-BPS determinant operators in $\mathcal{N}=4$ SYM at next-to-leading order at weak coupling. We use two complementary methods recently developed for a class of determinant three-point functions: one is based on Feynman diagrams and it extracts perturbative data at finite $N$, while the other one expresses a generic correlator of determinants as the zero-dimensional integral over an auxiliary matrix field. We generalize the latter approach to calculate one-loop corrections and we solve the four-point function in a semi-classical approach at large $N$. The results allow us to comment on the order of the phase transition that the four-point function is expected to exhibit in an exact integrability-based description.

\end{abstract}

\maketitle
\section{Introduction}
\label{secIntro}

Most of the advances in $\mathcal{N}=4$ super-Yang-Mills (SYM) theory and the gauge/gravity duality \cite{Maldacena:1997re} have been driven by the study of operators with bare dimension $\Delta$ small compared to $N$, especially single-trace operators. The interest has been raised by the connection to the integrability of SYM \cite{Minahan:2002ve} and the potential of paving the way to the complete solution of a non-trivial four-dimensional quantum field theory at finite coupling.
However, they encompass a limited sector of the theory: there are also ``heavy" operators whose dimensions scale as $\Delta\sim N$ \cite{Balasubramanian:2001nh} or $\Delta \sim N^2$ \cite{Lin:2004nb}. Correlators of such operators pose new challenges that make their study not a simple generalization of what was learned with single traces. Since they contain a number of fields that grows parametrically with $N$, there is no notion of planarity which would select the dominant diagrams at large $N$ \cite{Witten:1998xy} and traditional methods do not keep up with the factorial growth of the number of Feynman diagrams. Away from weak coupling, the status of integrability remains less developed and one does not know how it wholly manifests itself in the heavy sector.

In this work we focus on the class of 1/2-BPS determinant operators with $\Delta=N$ \cite{Balasubramanian:2001nh} and dual to maximal giant gravitons D3-branes in $AdS_5\times S^5$, wrapping $S^3\subset S^5$ and located at the center of $AdS_5$ \cite{McGreevy:2000cw}. One of the recent advances has concerned the correlator of two determinants and one non-BPS single trace \cite{Jiang:2019xdz,Jiang:2019zig}. Physically, it is the counterpart of three-point ``baryon-baryon-meson'' coupling of QCD and holographically it describes the absorption of a closed string by a giant graviton, which was the subject of numerous studies in the BPS sector \cite{Corley:2001zk,Bak:2011yy,Bissi:2011dc,Caputa:2012yj,Lin:2012ey,deMelloKoch:2019dda}. The progress in \cite{Jiang:2019xdz,Jiang:2019zig} was achieved in two steps: first mapping the three-point function to the overlap between an integrable boundary state (see also \cite{Jiang:2020sdw}) and a multi-particle state in the string worldsheet conformal field theory, then applying the idea of integrable bootstrap to derive an exact formula for such overlap in the framework of thermodynamic Bethe ansatz (TBA).
%First, the three-point function was identified with the overlap between an integrable boundary state (see also \cite{Jiang:2020sdw}), dual to the determinant pair, and a multi-particle state, dual to the single trace, in the string worldsheet CFT. Second, integrable-bootstrap techniques were used to derive an exact formula for such overlap in the framework of thermodynamic Bethe ansatz (TBA).

This result raises the question whether there are similar observables that are physically relevant and amenable to a TBA analysis. The four-point function of determinants is one of the most natural candidates \cite{Jiang:2019xdz}. It is expected to harbor a plethora of interesting physics \footnote{See sections 3.5.3 and 11.1 of \cite{Jiang:2019xdz} and references therein.}:  special (Regge and BFKL) kinematical limits and the Loschmidt echo of integrable spin chains, as well as the tachyonic instability of open-string modes and D-brane recombination on the AdS side. Moreover, it is believed to have a simple integrability description as the partition function of the worldsheet cylinder, whose ends are capped off by the boundary states of \cite{Jiang:2019xdz,Jiang:2019zig}.

In this paper we make the first preliminary steps toward such endeavor by collecting perturbative data on the four-point function
\begin{equation}
\label{eq5}
G_4
=
\langle \mathcal{D}_1 \mathcal{D}_2 \mathcal{D}_3 \mathcal{D}_4 \rangle
\end{equation}
of the determinant operators
\begin{flalign}
\label{eq26}
\mathcal{D}_i
&=
\det(y_i \cdot \Phi(x_i))\,,
\end{flalign}
where $y_i \cdot \Phi=\sum_{I=1}^6 y_i^I \Phi^I$ are linear combinations of the real scalar fields $\Phi^I$ in the adjoint representation of the gauge group $U(N)$ and the $y_i^I$ are six-dimensional null vectors ($y_i \cdot y_i=0$). The goal is to calculate quantum corrections up to subleading (one-loop) order at weak coupling. The usefulness of this data is twofold. First, it enables us to make basic, yet quantitative, remarks on the structure of the result, which should be embedded in any exact analysis. Second, it provides explicit data, especially in planar limit, to test the outcome of such analysis. The results also serve as a proof of concept for generating higher-loop data. In this paper we develop two methods introduced in \cite{Jiang:2019xdz}.

The first method combines standard perturbative techniques with a judicious strategy to solve the combinatorics of Wick contractions. Since determinants are defined by Levi-Civita symbols with $N$ indices (see \eqref{eq6} below), the evaluation of numerous tensor contractions is the main difficulty, although doable up to one loop. This leaves behind multiple sums of powers of the conformal cross ratios with a complicated dependence on $N$. This fact hinders the extrapolation to large $N$, except in a special limit of the vectors $y_i$ (see below \eqref{eq32}).

This observation leads to revisit the semi-classical approach of \cite{Jiang:2019xdz}. The strategy is unchanged: we cast the $m$-point function of determinants into a zero-dimensional integral over an auxiliary $m\times m$ matrix $\rho$, a form suitable for computing $1/N$-corrections in saddle-point approximation. Realizing an idea of \cite{Jiang:2019xdz}, we include the interacting Lagrangian perturbatively and express one-loop interactions as $\rho$-dependent insertions. In our case $m=4$, we discuss the saddle points of the (non-polynomial) $\rho$-action and their dominance as function of the cross ratios.
%, which basically corresponds to the three OPE expansion channels of the four-point function.
The extraction of a certain power of $1/N$ is only limited by the computational effort of expanding the action (as well as the insertions at one loop) to high order in the saddle-point fluctuations.

A further motivation behind the effective theory is in connection with the phase transitions that the cylinder partition function should exhibit at large $N$ \cite{Jiang:2019xdz}. This behavior is caused by the rapid growth of closed-string states when the cylinder reaches a critical length,
%This behavior should be due to the rapid growth of closed-string states when the cylinder reaches a critical length, or equivalently to tachyons in the open-string spectrum.
in analogy to the transition of the torus partition function
%confinement/deconfinement transition
at the Hagedorn temperature
% in QCD or pure Yang-Mills theory
\cite{Atick:1988si}. The cylinder length plays the role of the inverse temperature and it is controlled by the cross ratios. For our purposes, the effective theory is expected to detect the transition, at weak coupling at least, since the $\rho$-fields can be interpreted as open-string fields on the giant graviton D-branes and the transition separates different D-branes configurations \cite{Jiang:2019xdz}.
% made of two geodesic D-branes, each connecting a pair of determinant operators.
In this work we partially answer the question in \cite{Jiang:2019xdz} about the order of the transition: in free theory the phases are separated by a second-order transition at a ``Hagedorn'' cross ratio. However, our effective theory at one loop is affected by the choice of regularization scheme to deal with loop divergences, so it cannot discern whether the quantum behavior of the transition is truly the same or the phases become separated by a first-order transition, which generically occurs before reaching the ``Hagedorn'' cross ratio, in the interacting theory.

This paper is organized as follows. We begin with a summary of the results in section \ref{secR} and explain their derivation using the diagrammatical approach in section \ref{secPCGG} and the effective theory in section \ref{secET}. We make the remarks about the phase transition in section \ref{secPT} and outline perspectives on future work in section \ref{secConclusion}. The essential technical details are spelled out in the appendices.

\section{Results}
\label{secR}
We calculate the conformally-invariant correlator
\begin{equation}
\label{eq3}
\tilde{G}_4
=
\frac{G_4}
{\langle \mathcal{D}_1 \mathcal{D}_2 \rangle \langle \mathcal{D}_3 \mathcal{D}_4 \rangle}
= \tilde{G}_4^{(0)} + g^2 \tilde{G}_4^{(1)}+O(g^4)\,.
\end{equation}
We define $g^2
%=\lambda/(16\pi^2)
=g^2_{\textrm{YM}}N/(16\pi^2)$ and $g_{\textrm{YM}}$ is the SYM coupling.
The tree-level and one-loop orders are functions of the $SO(2,4)$ and $SO(6)$ cross ratios
\begingroup \allowdisplaybreaks 
\begin{gather}
\label{eq8}
z\bar{z}=\frac{x_{12}^2 x_{34}^2}{x_{13}^2 x_{24}^2}\,,
~~~~
(1-z)(1-\bar{z})=\frac{x_{14}^2 x_{23}^2}{x_{13}^2 x_{24}^2}\,,
\\
\label{eq9}
\alpha\bar{\alpha}=\frac{y_{12}y_{34}}{y_{13}y_{24}}\,,
~~~~
(1-\alpha)(1-\bar{\alpha})=\frac{y_{14}y_{23}}{y_{13}y_{24}}\,,
\end{gather}
\endgroup
where $x_{ij}^2=(x_i-x_j)^2$ and $y_{ij}=y_i \cdot y_j$,
and their combinations
\begin{gather}
\label{eq4}
u_1= \frac{z\bar{z}}{\alpha \bar{\alpha}}\,,
~~~~
u_2 =\frac{ (1-z)(1-\bar{z})}{(1-\alpha)(1-\bar{\alpha})}\,,
~~~~
r=\frac{u_1}{u_2}\,.
\end{gather}
%In this notation the normalization in \eqref{eq3} reads
%$\langle \mathcal{D}_1 \mathcal{D}_2 \rangle=N! \left(2g^2N^{-1}y_{12} \,x_{12}^{-2} \right)^N$.

At finite $N$, we derive the closed-form formula
\begin{flalign}
\label{eq24}
\tilde{G}_4^{(0)}&
=
\sum_{n=0}^{N}\sum_{m=0}^{N-n}c_{nm}\,u_{1}^{N-n}\,u_{2}^{n+m-N}
\end{flalign}
in terms of the coefficients
\begin{flalign}
\label{eq25}
c_{nm}&=\frac{1}{N!}\sum_{p,q,s}\frac{1}{\left(2n+p+q-N\right)!\,s!\,\left(q-p+s\right)!}
\\
&
\times\left[\frac{n!\,m!\,\left(N-n-m\right)!}{\left(m-p\right)!\left(N-n-m-q\right)!\left(p-s\right)!}\right]^{2}\,.
\nonumber
\end{flalign}
In this expression the multiple sum runs over the ranges
$p\in\left[0,m\right]$,
$q\in\left[\max\left(0,N-2n-p\right),N-n-m\right]$
and ~
$s\in\left[\max\left(0,p-q\right),p\right]$. We prove it with different methods in sections \ref{secPCGG} and \ref{secET}.

The large-$N$ asymptotics take a manifestly crossing-symmetric form at tree level ($n=0$) and one loop ($n=1$)
\begin{flalign}
\label{eq21}
\tilde{G}_4^{(n)}
&=f_{(n)}\left(u_{1},u_{2}\right)+f_{(n)}\left(1/u_{1},u_{2}/u_{1}\right)u_{1}^{N}
\\
&+f_{(n)}\left(u_{2},u_{1}\right)\left(\frac{u_{1}}{u_{2}}\right)^{N}
\nonumber
\end{flalign}
in terms of the $s$-channel functions
\begin{flalign}
\label{eq22}
&f_{(0)}\left(u_{1},u_{2}\right)=
\frac{u_{2}}{\left(1-u_{1}\right)\left(u_{2}-u_{1}\right)}
-\frac{2u_{1}^{2}u_{2}N^{-1}}{\left(1-u_{1}\right)^{2}\left(u_{2}-u_{1}\right)^{2}}
\nonumber
\\
&
\!\!\!+\frac{2u_{1}^{2}u_{2}\left(3u_{1}^{2}+u_{1}u_{2}+u_{1}+u_{2}\right)N^{-2}}{\left(1-u_{1}\right)^{3}\left(u_{2}-u_{1}\right)^{3}}+O\left(N^{-3}\right)\,,
\\
\label{eq23}
&
f_{(1)}\left(u_{1},u_{2}\right)=
-\frac{2\alpha^{2}\bar{\alpha}^{2}\left(z-\alpha\right)\left(z-\bar{\alpha}\right)\left(\bar{z}-\alpha\right)\left(\bar{z}-\bar{\alpha}\right)}{\left(z\bar{z}-\alpha\bar{\alpha}\right)^{2}}
\nonumber
\\
&
\times\frac{z\bar{z}\left(1-z\right)\left(1-\bar{z}\right)F^{\left(1\right)}\left(z,\bar{z}\right)}{\left[z\bar{z}\left(\alpha+\bar{\alpha}-1\right)-\alpha\bar{\alpha}\left(z+\bar{z}-1\right)\right]^{2}}+O\left(N^{-1}\right)
\,,
\end{flalign}
with the box master integral \cite{Usyukina:1992jd, Beisert:2002bb} (in the form of \cite{Jiang:2019xdz})
\begin{flalign}
\label{eq32}
\!
F^{\left(1\right)}\left(z,\bar{z}\right)&\!=\frac{2\,\textrm{Li}_{2}\left(z\right)-2\,\textrm{Li}_{2}\left(\bar{z}\right)+\log\left(z\bar{z}\right)\log\frac{1-z}{1-\bar{z}}}{z-\bar{z}}\,.
\end{flalign}
We prove them in section \ref{secET} under the assumptions ${u_1\neq 1}$, ${u_2\neq 1}$ and  ${u_1\neq u_2}$. For given $u_1$ and $u_2$, the formula \eqref{eq21} is to be understood as the only dominant term in it at large $N$. Further orders in $1/N$ can be derived effortlessly. The overall factor $\left(z-\alpha\right)\left(z-\bar{\alpha}\right)\left(\bar{z}-\alpha\right)\left(\bar{z}-\bar{\alpha}\right)$ in \eqref{eq23} is consistent with the superconformal Ward identities for four-point functions of 1/2-BPS operators \cite{Eden:2000bk, Chicherin:2015edu}. The presence of $F^{\left(1\right)}$ at one-loop order agrees with the expectation from conformal symmetry and lightcone OPE analysis \cite{Arutyunov:2003ae,Chicherin:2015edu}.

We also study a limit that forbids Wick contractions between determinants labeled by non-consecutive numbers in \eqref{eq5}. This corresponds to setting $y_{13}=y_{24}=0$ in \eqref{eq9}, which implies $u_1,u_2\to 0$ with their ratio $r$ finite in \eqref{eq4}. We obtain for any $N$ and $r$
\begin{flalign}
\label{eq1}
&\tilde{G}_4^{(0)}|_{y_{13}=y_{24}=0}=\frac{1-r^{N+1}}{1-r}\,,
\\
\label{eq2}
&\tilde{G}_4^{(1)}|_{y_{13}=y_{24}=0}
=
\frac{2\left(N+1\right)r}{N^{2}\left(1-r\right)^{3}}
\, F^{\left(1\right)}\left(z,\bar{z}\right)
\\
&
~~\times
\left[N\left(r-1\right)+r+1+r^{N}\left(N\left(r-1\right)-r-1\right)\right]\,.
\nonumber
\end{flalign}
We derive them in section \ref{secPCGG} and check that the large-$N$ analysis is consistent with them in section \ref{secET}.

We checked that our results cannot be derived from the general one-loop expressions for 1/2-BPS operators of equal weights $k$ in \cite{Arutyunov:2003ae, Chicherin:2015edu}, which follow from arguments based on symmetries and singularity structure for $k \ll N$ and $N\gg 1$. Determinants operators, with $k=N$, fail to satisfy this condition. One can appreciate the importance of this fact in gauge theory \cite{Balasubramanian:2001nh}: after expressing determinants as multi-trace operators of length $N$, non-planar diagrams survive the planar limit because they carry combinatorial factors that overwhelm the suppression by powers of $1/N^2$. Therefore one cannot project each determinant onto single-trace contributions in calculating \eqref{eq5}, which prevents us from making contact with the analogous correlator of 1/2-BPS single traces. We comment further in section \ref{secConclusion}.

\section{Diagrammatic approach}
\label{secPCGG}
We give a brief account of the standard perturbative calculation adapted to heavy operators \cite{Berenstein:2003ah}.
%In the absence of a useful notion of planarity which would select the dominant Feynman diagrams at large $N$, the results naturally come out as functions of finite $N$. Combined with some results at one loop in \cite{Jiang:2019xdz},
%... this method becomes ideal for checking the results obtained in the approach of section \ref{secET}.
The key object is an operator of length $2L$, called ``partially-contracted giant graviton" $\mathcal{G}_{L}$ in \cite{Jiang:2019xdz}, which stems from the Wick contractions between two determinants, say $\mathcal{D}_1$ and $\mathcal{D}_2$:
\begin{flalign}
\label{eq27}
\sum_{L=0}^{N}\Delta_{12}^{N-L}\,\mathcal{G}_{L}(x_1,x_2)\,.
\end{flalign}
We conveniently factor out the scalar propagators ${\Delta_{ij}=g_{\textrm{YM}}^2 I_{x_{i}x_{j}}y_{ij}/2}$ due to the Wick contractions (with the abbreviation $I_{xy}=(4\pi^2 (x-y)^2)^{-1}$) and define the tree-level $O(g_{\textrm{YM}}^0)$ and one-loop $O(g_{\textrm{YM}}^2)$ parts in the expansion 
$
\mathcal{G}_{L}(x_1,x_2)=
\mathcal{G}_{L}^{\left(0\right)}(x_1,x_2) + \, \mathcal{G}_{L}^{\left(1\right)}(x_1,x_2) + \dots
$ at weak coupling (see appendix \ref{secPCGGdetails}). The intermediate step of contracting determinants in pairs is not as crucial as for correlators involving single traces \cite{Jiang:2019xdz}, but it remains a useful object for building some diagrams at one loop.
\paragraph{Tree level}
The strategy is to write \eqref{eq26} as a product of scalar fields with the $U(N)$-indices contracted with Levi-Civita symbols:
\begin{gather}
\label{eq6}
\mathcal{D}_i
=
\frac{\varepsilon_{i_{1}...i_{N}}\varepsilon^{j_{1}...j_{N}}}{N!}
\left(y_{i}\cdot\Phi_{\phantom{i_{1}}j_{1}}^{i_{1}}\right)
\dots\left(y_{i}\cdot\Phi_{\phantom{i_{1}}j_{N}}^{i_{N}}\right)\,.
\end{gather}
The diagrams are specified by the numbers of propagators between pairs of determinants (the bridge lengths $\ell_{ij}$). It is simple to express $\ell_{14}=\ell_{23}=N-\ell_{12}-\ell_{13}$,
$\ell_{24}=\ell_{13}$
and
$\ell_{34}=\ell_{12}$
in terms of $\ell_{12}$ and $\ell_{13}$. The Wick contractions produce a complicated structure of $\epsilon$-symbols, which we cast into a combinatorial problem
%and solve at fixed $\ell_{12}$ and $\ell_{13}$
in appendix \ref{secPCGGdetails}. The solution is basically the coefficient \eqref{eq25}, where $n$ and $m$ are a relabeling of $\ell_{12}$ and $\ell_{13}$ respectively. The sum over all values of the bridge lengths yields \eqref{eq24}.
%that for single-trace operators due to $\epsilon$-symbols in \eqref{eq6}

We were not able to write \eqref{eq25} as a simpler function of $N$. However, this task becomes trivial if $m=0$.  This contribution dominates the sum over $m$ in \eqref{eq24} in the limit described above \eqref{eq1}. The remaining sum over $n$ is geometric and gives \eqref{eq1}.

\paragraph{One loop} We move on to the subleading order under the assumption $y_{13}=y_{24}=0$ which allows for an exact summation over the bridge lengths. The basic one-loop interactions are three: the scalar self-energy (a), the scalar quartic interaction (b) and the gluon-exchange interaction (c). The expressions are known in literature in terms of the integrals $Y_{x_1,x_2,x_3}$, $X_{x_1,x_2,x_3,x_4}$ and $F_{x_1,x_2,x_3,x_4}$ \cite{Beisert:2002bb, Drukker:2008pi, Drukker:2009sf} (see also appendix \ref{secConventions}). The singular limits, when two points are coincident, can be regularized introducing a spacetime cutoff $\epsilon\ll 1$ in point-splitting regularization. The interactions carry a non-trivial $U(N)$-index structure and the contractions of $\epsilon$-symbols are much harder than that at tree level.
%In order to approach this step systematically,  

We organize the diagrams into three classes.
%, depending on the number of determinants that the insertions involve.
Interactions can connect the scalars of two determinants (e.g. one scalar in $\mathcal{D}_1$ and one in $\mathcal{D}_2$ for type (a), or two scalars in $\mathcal{D}_1$ and two in $\mathcal{D}_2$ for types (b) and (c)). The corresponding diagrams combine into
\begin{flalign}
\label{eq28}
& -g_{\textrm{YM}}^2\frac{\left(N!\right)^{2}\left(N+1\right)}{N}\sum_{\ell_{13}=1}^{N-1}\ell_{13}\left(N-\ell_{13}\right)
\left(\Delta_{12}\Delta_{34}\right)^{N-\ell_{13}}
\nonumber
\\
& \times \left(\Delta_{14}\Delta_{23}\right)^{\ell_{13}}\frac{Y_{x_1x_1x_2}+Y_{x_1x_2x_2}}{I_{x_1x_2}}+\textrm{perm.}\,,
\end{flalign}
where ``perm." stands for the cyclic shifts of the indices $\{1,2,3,4\}$.
%Interactions can be of type (b)-(c) when connecting more than two determinants.
Diagrams with three-determinant interactions (e.g. connecting one scalar in $\mathcal{D}_1$, two in $\mathcal{D}_2$ and one in $\mathcal{D}_3$) give
\begingroup \allowdisplaybreaks 
\begin{flalign}
\label{eq29}
&g_{\textrm{YM}}^2\frac{\left(N!\right)^{2}\left(N+1\right)}{N}\sum_{\ell_{12}=0}^{N-2}
\left(\ell_{12}+1\right)
\\
&
\times
\left(N-1-\ell_{12}\right)
\left(\Delta_{12}\Delta_{34}\right)^{\ell_{12}+1}
\left(\Delta_{14}\Delta_{23}\right)^{N-1-\ell_{12}}
\nonumber
\\
&
\left(\frac{X_{x_1x_2x_2x_3}}{I_{x_1x_2}I_{x_2x_3}}-F_{x_1x_2,x_2x_3}\right)+\textrm{perm.}\,,
\nonumber
\end{flalign}
\endgroup
while those with four-determinant interactions (connecting four scalars, one in each determinant) give
\begin{flalign}
\label{eq30}
&g_{\textrm{YM}}^2\frac{\left(N!\right)^{2}\left(N+1\right)}{N}\sum_{\ell_{12}=0}^{N-1}\left(\Delta_{12}\Delta_{34}\right)^{\ell_{12}}\left(\Delta_{14}\Delta_{23}\right)^{N-1-\ell_{12}}
\nonumber
\\
& \times \left\{ \left(-1-2\ell_{12}+\ell_{12}^{2}+N-\ell_{12}N\right)\Delta_{12}\Delta_{34}\frac{X_{x_1x_2x_3x_4}}{I_{x_1x_2}I_{x_3x_4}}\right.
\nonumber
\\
&
\phantom{\times}+\left(2+4\ell_{12}+\ell_{12}^{2}-2N-\ell_{12}N\right)\Delta_{14}\Delta_{23}\frac{X_{x_1x_2x_3x_4}}{I_{x_1x_4}I_{x_2x_3}}
\nonumber
\\
&
\phantom{\times}-\left(\ell_{12}+1\right)\left(N-1-\ell_{12}\right)\Delta_{12}\Delta_{34}F_{x_1x_2,x_3x_4}
\\
&
\left.\phantom{\times}+\ell_{12}\left(N-\ell_{12}\right)\Delta_{14}\Delta_{23}F_{14,23}
\vphantom{\frac{X_{1234}}{I_{12}I_{34}}}
\right\} \,.
\nonumber
\end{flalign}
The sum of \eqref{eq28}-\eqref{eq30} yields \eqref{eq2}. The terms containing $X_{x_1x_2x_3x_4}$, which is proportional to the conformal integral $F^{(1)}(z,\bar{z})$ in \eqref{eq32}, produce the full result, whereas all other terms, which depend on the cutoff and the non-conformal integrals $Y$'s, cancel at each value of $\ell_{12}$.

\section{Effective theory approach}
\label{secET}
In this section we generalize the semi-classical approach, constructed in free theory in \cite{Jiang:2019xdz}, to calculate the one-loop correction of the $m$-point function
\footnote{One can also include single traces without significant changes. In the step 2 below, they would multiply the traces that come from the expansion of the interacting action $S_{\textrm{YM}}^{\textrm{(int)}}$ in \eqref{eq33}. The integral over SYM fields would produce also Wick contractions among the extra traces and between these and the interaction vertices.}
\begin{equation}
\label{eq10}
G_{m}=\left\langle \mathcal{D}_{1}\dots\mathcal{D}_{m}\right\rangle \,.
\end{equation}
We closely follow section 3.1 therein and highlight the changes needed to accommodate one-loop corrections in the final formula \eqref{eq11}-\eqref{eq12} below.
\paragraph{Step 1}
We express the determinants as integrals over zero-dimensional fermions $\chi^{a}_k$ (and their conjugates $\bar{\chi}_{k,a}$). The index $a=1,\dots, N$ is in the (anti-)fundamental representation of $U(N)$ and $k=1,\dots, m$ labels the fermions.
\begin{flalign}
\label{eq14}
G_{m}&=\frac{1}{Z}\int DA_\mu D\Phi^I D\Psi Dc\, D\bar{c}\, d\chi_k d\bar{\chi}_k\,e^{-S_{\textrm{YM}}(A_\mu,\Phi^I,\Psi,c)}
\nonumber
\\
& \times e^{\sum_{k=1}^{m}\bar{\chi}_{k,a}\left(y_{k}\cdot\Phi\left(x_k\right)\right)^a_{~b}\chi_{k}^b}
\end{flalign}
The path integral is over the gluons $A_\mu$, the scalars $\Phi^I$, the gluinos $\Psi$ and the ghosts $c$, $\bar{c}$. It is weighted by the full action of the theory $S_{\textrm{YM}}$ (in the form of \cite{Erickson:1999qv,Erickson:2000af}, see appendix \ref{secConventions}) and the corresponding partition function is $Z$. We suppress $U(N)$-indices and abide by a positional notation: $\bar{\chi}_k\chi_l=\bar{\chi}_{k,a}\chi^{a}_l$ is a singlet, $\chi_k \bar{\chi}_l=\chi^{a}_k\bar{\chi}_{l,b}$ is a matrix and $\bar{\chi}_k \Phi^I \chi_l=\bar{\chi}_{k,a} (\Phi^I)^a_{~b} \chi^{b}_l$ is a bilinear.

\paragraph{Step 2} We integrate out $\Phi^I$ in two sub-steps. First, we split the action into the quadratic (free) and interacting (int) parts and complete the square in $\Phi^I$:
\begin{flalign}
\label{eq33}
G_{m}&=\frac{1}{Z}\int DA_\mu D\Phi^I D\Psi Dc\, D\bar{c}\, d\chi_k d\bar{\chi}_k\,
e^{-S_{\textrm{YM}}^{\left(\textrm{int}\right)}\left(A_\mu,\Phi^I,\Psi,c\right)}
\nonumber
\\
&~~~~ \times 
e^{-S_{\textrm{YM}}^{\left(\textrm{free}\right)}\left(A_\mu,\Phi^I-S^I,\Psi,c\right)
-g_{\textrm{YM}}^{-2} \int d^4x\, \textrm{tr}\left(S^{I}\square S_{I}\right)}
\end{flalign}
The last term in the exponent compensates for the shift of the scalars by $S^{I}\left(x\right)=-g_{\textrm{YM}}^{2}\sum_{k=1}^m I_{xx_{k}}\,y_{k}^{I}\chi_{k}\,\bar{\chi}_{k}/2$ (with $I_{xy}$ defined below \eqref{eq27}) in the quadratic action $S_{\textrm{YM}}^{\left(\textrm{free}\right)}\left(A_\mu,\Phi^i,\Psi,c\right)=-g_{\textrm{YM}}^{-2} \int d^4x\, \textrm{tr}\left(\Phi^I \square \Phi^I\right)+\dots$.

Next, we change variables $\Phi^I=\Phi'^I+S^I$ and perform the path integral in $\Phi'^I$ and the SYM fields perturbatively. In order to do so, we bring down vertices from the interacting action and consider a correlator with insertions of multiple single-trace operators. This mimics the construction of the interactions (a)-(c) in section \ref{secPCGG}, with the difference that they are built from an action with a ``classical background'' $S^I$ for $\Phi'^I$.

We are now ready to path integrate. There would be actually no integration in free theory, because one can drop $\Phi'^I$ from the integrand and ignore the other SYM fields. When the coupling is turned on, one has instead to carefully perform free Wick contractions and keep diagrams with one $S^I$ at least, which do not cancel against the normalization $1/Z$. We did so for the simpler diagrams \eqref{eq19} and \eqref{eq20} below, but it is very lengthy already at one loop, so we devise a shortcut in appendix \ref{secETdetails}. The outcome of the integration mimics the structure of the basic interactions (a)-(c) in section \ref{secPCGG}
\begin{gather}
\label{eq16}
\mathcal{O}_{\textrm{1-loop}}^{S}
=
\mathcal{O}_{a}^{S}+\mathcal{O}_{b}^{S}+\mathcal{O}_{c}^{S}\,,
\end{gather}
with the scalar self-energy
\begin{flalign}
\label{eq18}
&
 \mathcal{O}_{a}^{S}=
-\frac{g_{\textrm{YM}}^{4}N}{4}\sum_{k_{1}\neq k_{2}}\left(Y_{x_{k_{1}}x_{k_{1}}x_{k_{2}}}+Y_{x_{k_{1}}x_{k_{2}}x_{k_{2}}}\right)
\\
&
\!\!
\times y_{k_{1}k_{2}}\left[\textrm{tr}\left((\chi_{k_{1}}\bar{\chi}_{k_{1}})(\chi_{k_{2}}\bar{\chi}_{k_{2}})\right)-\frac{\textrm{tr}\left(\chi_{k_{1}}\bar{\chi}_{k_{1}}\right)\textrm{tr}\left(\chi_{k_{2}}\bar{\chi}_{k_{2}}\right)}{N}\right]\,,
\nonumber
\end{flalign}
the scalar quartic interaction
\begin{flalign}
\label{eq19}
\mathcal{O}_{b}^{S}&=
-\frac{g_{\textrm{YM}}^{6}}{16}\sum_{k_{1}\neq k_{2}}\sum_{k_{3}\neq k_{4}}X_{{x_{k_{1}}x_{k_{2}}x_{k_{3}}x_{k_{4}}}}
y_{k_{1}k_{2}}
\\
&
\times y_{k_{3}k_{4}} \textrm{tr}\left((\chi_{k_{1}}\bar{\chi}_{k_{1}})(\chi_{k_{3}}\bar{\chi}_{k_{3}})\left[\chi_{k_{4}}\bar{\chi}_{k_{4}},\chi_{k_{2}}\bar{\chi}_{k_{2}}\right]\right)\,,
\nonumber
\end{flalign}
and the gluon-exchange interaction
\begin{flalign}
\label{eq20}
\mathcal{O}_{c}^{S}&=
-\frac{g_{\textrm{YM}}^{6}}{16}\sum_{k_{1}\neq k_{2}}\sum_{k_{3}\neq k_{4}}I_{x_{k_{1}}x_{k_{2}}}I_{k_{3}k_{4}}F_{{x_{k_{1}}x_{k_{2}}x_{k_{3}}x_{k_{4}}}}
\\
&
\times 
y_{k_{1}k_{2}} y_{k_{3}k_{4}}
\textrm{tr}\left((\chi_{k_{1}}\bar{\chi}_{k_{1}})(\chi_{k_{2}}\bar{\chi}_{k_{2}})(\chi_{k_{3}}\bar{\chi}_{k_{3}})(\chi_{k_{4}}\bar{\chi}_{k_{4}})\right)\,.
\nonumber
\end{flalign}
The symbol $I_{xy}$ is defined below \eqref{eq27} and the integrals $Y$, $X$ and $F$ are mentioned above \eqref{eq28}.

In summary, step 2 leads to
\begin{flalign}
\label{eq15}
G_{m}&=\int d\chi_k \, d\bar{\chi}_k\left(1+\mathcal{O}_{\textrm{1-loop}}^{S}+\dots\right)
\\
&
\times
e^{-\frac{g^{2}}{N}\sum_{k\neq l}d_{kl}\left(\bar{\chi}_{k}\chi_{l}\right)\left(\bar{\chi}_{l}\chi_{k}\right)}\,.
\nonumber
\end{flalign}
The quartic action in fermions comes from $S^I \square S^I$ in \eqref{eq33} and $d_{kl}=y_{kl}/x_{kl}^2$. The difference with free theory is the insertion of an effective one-loop operator, as well as higher-order ones omitted in the dots. The quartic action is  unchanged as much as the rest of the derivation.

\paragraph{Step 3} We linearise the quartic fermions in \eqref{eq15} via a Hubbard-Stratonovich transformation
\begin{flalign}
G_{m}&=\frac{1}{Z_{\rho}}\int d\chi_k \, d\bar{\chi}_k \,d\rho \,\left(1+\mathcal{O}_{\textrm{1-loop}}^{S}+\dots\right)
\\
&
\times
e^{-\frac{N}{g^{2}}\sum_{k,l=1}^m\rho_{kl}\rho_{lk}+2i \sum_{k,l=1}^m\hat{\rho}_{kl}\bar{\chi}_{l}\chi_{k}}\,,
\nonumber
\end{flalign}
with
\begingroup \allowdisplaybreaks 
\begin{gather}
\hat{\rho}_{kl}=\sqrt{d_{kl}}\,\rho_{kl}\,, ~~~ Z_{\rho}=\left(\frac{g^{2}\pi}{2N}\right)^{m\left(m-1\right)/2}\,,
\\
d\rho=\prod_{k<l}d\textrm{Re}\left(\rho_{kl}\right)\,d\textrm{Im}\left(\rho_{kl}\right)\,.
\end{gather}
\endgroup
This introduces the integral over the traceless $m\times m$ matrix $\rho_{kl}$. Minor sign changes from \cite{Jiang:2019xdz} ensure $\rho^\dagger=\rho$ and make superfluous the discussion of reality conditions and convergence on a case-by-case basis, like in section 3.5 therein.
%$d\rho = \prod_{i<j}d\textrm{Re}\left(\rho_{ij}\right)\,d\textrm{Im}\left(\rho_{ij}\right)$

\paragraph{Step 4 and result} We integrate out fermions:
\begin{equation}
\label{eq11}
G_{m}=\frac{1}{Z_{\rho}}\int d\rho\left(1+\left\langle \mathcal{O}_{\textrm{1-loop}}^{S}\right\rangle _{\chi}+\dots\right)\,e^{-NS_{\textrm{eff}}\left[\rho\right]}\,.
\end{equation}
The action for the auxiliary matrix
\begin{flalign}
\label{eq13}
S_{\textrm{eff}}\left[\rho\right]&=\frac{1}{g^{2}}\textrm{tr}\left(\rho^2\right)-\log\textrm{det}\left(-2i\hat{\rho}\right)
\end{flalign}
comes with a power of $N$ in \eqref{eq11}, making this formula an ideal starting point for extracting $1/N$-corrections in saddle-point expansion. The one-loop insertion is averaged over fermions
\begin{flalign}
\label{eq12}
\left\langle \mathcal{O}^{S}_{\textrm{1-loop}}\right\rangle _{\chi}&=\frac{\int d\chi_k \, d\bar{\chi}_k \, \mathcal{O}^{S}_{\textrm{1-loop}}\,e^{2i\sum_{k,l=1}^m\hat{\rho}_{kl}\,\bar{\chi}_{l,a}\chi_{k}^a}}{\int d\chi_k \, d\bar{\chi}_k\,e^{2i\sum_{k,l=1}^m\hat{\rho}_{kl}\,\bar{\chi}_{l,a}\chi_{k}^a}}
\end{flalign}
and it can be computed for fixed $\rho$ performing Wick contractions on \eqref{eq18}-\eqref{eq20} with the Feynman rule
$\left\langle \bar{\chi}_{k,a}\chi_{l}^{b}\right\rangle _{\chi}=-\frac{i}{2}\delta_{a}^{b}\left(\hat{\rho}^{-1}\right)_{kl}$. Notice that the fermionic operators are formally divergent (see above \eqref{eq28}). In fact, divergences can only cancel out after the $\rho$-integration of the basic interactions $\langle\mathcal{O}_{a}^{S}\rangle_\chi$, $\langle\mathcal{O}_{b}^{S}\rangle_\chi$ and $\langle\mathcal{O}_{c}^{S}\rangle_\chi$, as much as it happens after ``inserting'' their counterparts (a)-(c) in section \ref{secPCGG} in all possible ways into the diagrams \eqref{eq28}-\eqref{eq30} (see below \eqref{eq30}).

\paragraph{Finite-$N$ analysis}
The study of the case $m=4$ is the main application of the effective theory in this paper. Ignoring the presence of \eqref{eq12} in \eqref{eq11}, we can bring down the determinant in \eqref{eq13} from the exponent and expand its $N$-th power with the help of the multinomial theorem. The resulting integral of a polynomial in $\rho_{kl}$ times a Gaussian weight (given by the first term in \eqref{eq13}) reproduces \eqref{eq24}-\eqref{eq25}. The multiple sums that appear here come from the combinatorics of Wick contractions in the diagrammatic approach, or from the use of the multinomial theorem and the $\rho$-integration in the effective theory. This suggests that Wick contractions correspond to $\rho$-integration, which corroborates the expectation below \eqref{eq12} about the mechanism of divergence cancellation at loop level.

\paragraph{Large-$N$ analysis}
The evaluation of the full integral \eqref{eq11} is not possible exactly, so we perform a saddle-point expansion for large $N$. The action \eqref{eq13} has three saddle points $\rho^*$ for generic values of $y_{ij}$ \footnote{\label{foot1} We find convenient to study the saddle-point equations $\rho_{kl}=\frac{g^{2}}{2}\left(\hat{\rho}^{-1}\right)_{kl}\sqrt{d_{kl}}$ with $k,l=1,\dots, 4$ in polar coordinates for the $\rho_{kl}$.
%In particular, we complete the analysis in \cite{Jiang:2019xdz} to verify that I-III are the only solutions.
The space of solutions is enhanced for special values of the cross ratios \eqref{eq4}, e.g. $r=1$ in the case $y_{13}=y_{24}=0$.
%the values $u_1=1$, $u_2=1$ or $u_1=u_2$.
We do not study the semi-classical fluctuations in these cases (see below \eqref{eq32}).}.
The first solution (I) is
\begin{gather}
\label{eq17}
\rho^*_{(I)}:
~~~~
\left|\rho_{12}\right|^{2}=\left|\rho_{34}\right|^{2}=\frac{g^{2}}{2}\,,
~~~~
\textrm{other  } \rho_{kl}=0
\,,
\end{gather}
while the second (II) is obtained from it by swapping the indices $2\leftrightarrow 3$ and the third (III) with $2\leftrightarrow 4$. We extend the discussion in section 3.5.3 of \cite{Jiang:2019xdz} to quantify the fluctuations around the solutions. The procedure is fairly standard and we only sketch it concisely.

The solution I is favored (namely $S_{\textrm{eff}}[\rho^*_{(I)}]<S_{\textrm{eff}}[\rho^*_{(i)}]$ with $i=$\,\,II, III) for $u_1<1$ and $u_1<u_2$. It has two flat directions corresponding to the complex phases of $\rho_{12}$ and $\rho_{34}$, therefore we express $\rho_{12}=\left({g}/{\sqrt2}+r_{12}\right)e^{i \theta_{12}}$ and $\rho_{34}=\left({g}/{\sqrt2}+r_{34}\right)e^{i \theta_{34}}$ and take the fluctuations ($r_{12}$, $r_{34}$ and the 4 remaining complex $\rho_{ij}$) to be of order $N^{-1/2}$. We can then expand the integrand in \eqref{eq11}, evaluate the Gaussian integrals and integrate exactly over the moduli $\theta_{12}$ and $\theta_{34}$. The result gives the first term in \eqref{eq21} with the explicit functions \eqref{eq22} and \eqref{eq23}. We also check that pieces proportional to $\log\epsilon$ and $Y$ vanish up to the order of $1/N$ in these formulas. The analysis for the solutions II and III can be inferred from crossing symmetry and it gives the second and third term in \eqref{eq21} respectively.

We conclude the section with a close look to the case $y_{13}=y_{24}=0$. The solutions I and III dominate for $r<1$ and for $r>1$ respectively. The sum of the two contributions delivers \eqref{eq1} and \eqref{eq2}, where the terms with(out) $r^N$ arise from the fluctuations around III (I). The novelty of this case is that the series of $1/N$-corrections truncates, so one can actually regard \eqref{eq1} and \eqref{eq2} as exact functions in $N$, despite the fact that we assumed $N$ to be large in the derivation. The large-$N$ analysis would not rule out the appearance of exponential corrections $e^{-N}$ when $N$ becomes finite, but these are expected to appear in individual correlators of determinants, not in the normalized correlator \eqref{eq3}, as discussed in section 1.2 of \cite{Jiang:2019xdz}. Moreover, the result has a smooth limit for $r\to 1$, so one can also drop the assumption $r\neq 1$. In particular, we observe the scaling $\tilde{G}_4^{(0)}\sim N$ and $\tilde{G}_4^{(1)}\sim N^2$ when $r=1$, which is consistent with the presence of extra flat directions in the saddle point.

\section{Comments on phase transitions}
\label{secPT}

\begin{figure}[t]
\centering
\includegraphics[scale=0.5]{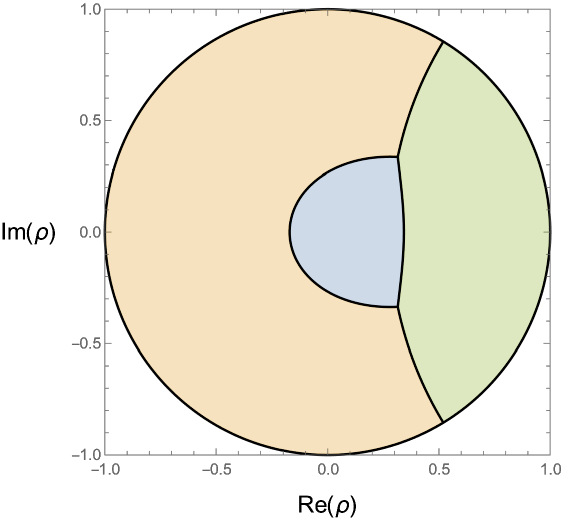}
\caption{Saddle points of the effective theory at zero coupling. The diagram is plotted in the parameter $\rho=\rho(z)$ of \cite{Hogervorst:2013sma} and fixed $\alpha \bar{\alpha}=10(1-\alpha)(1-\alpha)=1$. The region I (blue) is defined by $u_1,u_1/u_2<1$, the region II (yellow) by $1,u_1/u_2<u_1$ and the region III (green) by $1,u_1<u_1/u_2$. The region II vanishes in the limit $y_{13}=y_{24}=0$ (see above \eqref{eq1}). Each saddle point in \eqref{eq17} and below saturates \eqref{eq11} in the homonymous region. The boundary curves (black) separating the regions are the locus of critical points $z=z_c$ in section \ref{secPT}. Small values of $\rho\bar{\rho}$ favour the saddle point I, while the complex phase of $\rho$ plays a role to favour I, II or III at higher values.}
\label{fig}
\end{figure}

The study of the effective theory allows to put together quantitative remarks on the phase transitions of the four-point function (see section \ref{secIntro}).

We begin with the theory at $g=0$. Let us make a continuous variation of $z$ (and of $\bar{z}=z^*$ in Euclidean spacetime) that traces a curve crossing the boundary between two regions in figure \ref{fig}, for example I and II \footnote{One could also vary $\alpha$ and $\bar{\alpha}$ with the effect of reshaping the regions in figure \ref{fig}.}. The correlator exhibits one or more ``phase transitions'': when two saddle points contribute with equal weight ($S_{\textrm{eff}}[\rho^*_{(I)}]=S_{\textrm{eff}}[\rho^*_{(II)}]$ for a critical value $z=z_c$) or when the dominant saddle point becomes unstable because some masses undergo a sign change (at the ``Hagedorn'' cross ratio $z=z_H$). The transition is first order if $z$ takes on the value $z_c$ first and then $z_H$, or second order if $z_c=z_H$.

For a given variation of $z$, the critical value ${z=z_c}$ sits at the crossing point between two regions, where two elements in the set $\{1, u_1, u_1/u_2\}$ are equal. The effective theory is sensitive to $z_c$: the tree-level and one-loop data \eqref{eq21} discriminate three cases, depending on which element has the greatest value. The same remarks hold true when $y_{13}=y_{24}=0$: both the diagrammatic and effective-theory approaches are able to tell apart $r<1$ and $r>1$ in \eqref{eq1}-\eqref{eq2}. Based on these observations, we argue that $z_c$ depends only on the geometric information encoded into the ``cross ratios'' \eqref{eq4}, hence $z_c$ is independent of the coupling $g$.

This contrasts with $z_H=z_H(g)$, a function yet to be determined and defined by the emergence of tachyons in the open-string spectrum of the TBA partition function. Here we approach the determination of $z_H(g)$ by studying the stability of the saddle points of the effective theory. We analyze the spectrum of small fluctuations around the solution I without loss in generality. Denoting the vector of 10 real fluctuations (listed below \eqref{eq17}) with $\tilde{\rho}$, the effective action has the expansion $S_{\textrm{eff}}[\rho]= -2Ng^{-2}\tilde{\rho}^T M \tilde{\rho}+\dots $, where the mass matrix $M$ has doubly-degenerate eigenvalues
\begin{gather}
1\pm\sqrt{u_1}
\,,~~~
1\pm\sqrt{{u_1}/{u_2}}
\,,~~~
2\,.
\end{gather}
When the saddle point is dominant (region I in figure \ref{fig}), it is also stable because all masses are positive. In concomitance with a crossover (with II when $u_1=1$ or with III when $u_1=u_2$), some fluctuations become zero-modes. This indicates that $z_H(g)=z_c+O(g^2)$: the transition is second order in free theory and it occurs precisely at the ``Hagedorn'' cross ratio.

Turning on the coupling changes this situation, but the current formulation of the effective theory does not seem capable of addressing this question rigorously. Indeed, loop corrections do not enter \eqref{eq11} as corrections to the effective action, so they do not change the saddle points nor the fluctuation spectra. The structure of \eqref{eq11} is a direct consequence of our derivation after \eqref{eq33}, when we bring down interaction vertices from the SYM action as usual in perturbative calculations. One could be tempted to bring the one-loop operator in \eqref{eq11} back to the exponent and repeat the saddle-point analysis for $S_{\textrm{eff}}\left[\rho\right]-N^{-1} \langle \mathcal{O}_{\textrm{1-loop}}^{S}\rangle _{\chi}$. However, such new action would be unsatisfactory because $\eqref{eq16}$ depends on the regularization choice. The ambiguity is expected to disappear after the $\rho$-integration (see below \eqref{eq12}) because the four-point function is finite, but it would undermine any physical statement made before the $\rho$-integration, such as the determination of $r_H(g)$ from the fluctuation spectra. A workaround could prescribe to pick point-splitting regulatization and exponentiate only the terms proportional to $F^{(1)}\left(z,\bar{z}\right)$ in $\langle \mathcal{O}_{\textrm{1-loop}}^{S}\rangle _{\chi}$, which are known to reproduce the full result \eqref{eq2}. We refrain from pursuing this direction and add further comments in section \ref{secConclusion}.

\section{Conclusion}
\label{secConclusion}

In this paper we performed the weak-coupling computation of the four-point function of determinant operators in $\mathcal{N}=4$ SYM from two complementary angles. We adapted diagrammatical techniques to operators of large length, collected data at finite $N$ and identified a choice of the operators that reduces the results to a simple function of the cross ratios. We also revisited a semi-classical method for efficiently extracting $1/N$-corrections from correlators involving an arbitrary number of determinants. We explained how to include interactions at weak coupling, worked out an explicit formula for the one-loop order and extracted data for the four-point function, both at large $N$ and at finite $N$ at zero coupling. 
%The results of the two approaches are in perfect agreement.
The semi-classical method also shows that the expected phase transition of the four-point function is second order at zero coupling, but it lies precisely at the border between first and second order behavior.

This finding warrants further study. It would be interesting to devise an alternative form of the effective theory beyond tree level in order to determine the order of the transition. One could take a different path in step 2 of section \ref{secET}, where loop corrections force a choice of regularization scheme at an early stage, perhaps introducing an auxiliary field analogous to the fermion bilinear in the SYK model \footnote{A. Kitaev, talks at KITP on April 7 and May 27, 2015, \href{http://online.kitp.ucsb.edu/online/entangled15/}{http://online.kitp.ucsb.edu/online/entangled15/}.}. One could also explore the Regge and BFKL limits of the (Lorentzian) effective theory and study its (complex) action as in \cite{Copetti:2020dil}. Likewise, it would be interesting to attack the problem from the relevant classical D-brane configuration at strong coupling.

One could also extend the perturbative toolkit to higher-loop order, possibly in the elegant language of the Lagrangian insertion approach \cite{Chicherin:2015edu} (see end of section \ref{secR}) \footnote{In \cite{Chicherin:2015edu} the dominance of planar graphs is crucial to restrict the ansatz of the integrand of the four-point functions and it enables to set up a recursion relation based on the single traces of lowest weight. In order to extend this approach, one would have to generalize the ``universality'' property of the OPE structure constants and reconsider the recursion seed, perhaps studying determinants together with the closely-related subdeterminant operators \cite{Balasubramanian:2001nh}. One could also try to preliminarily operate free Wick contractions among the four determinants and consider the resulting effective operator (akin to \eqref{eq27} for a determinant pair) as the starting point of the analysis. The effective operator is non-local (depending on $x_1, \dots, x_4$) and perhaps a sum of multi traces, although only a single trace could suffice for bootstrapping the one-loop results in the present paper.}. Moving to other models, the analysis should be doable for heavy operators in SYM \cite{Chen:2019gsb}, ABJ(M) \cite{Chen:2019kgc}, the conformal fishnet theory \cite{Gurdogan:2015csr} and other chiral theories \cite{Caetano:2016ydc, Pittelli:2019ceq} with simple interaction Lagrangians. A study of loop corrections could unveil traces of integrability for new observables in less symmetric or prototypical models of broad interest.

\paragraph*{Acknowledgement}
We thank Yunfeng Jiang, Shota Komatsu and Matthias Wilhelm for useful discussions, in particular the first two also for collaboration on related topics. We are grateful to Shota Komatsu for comments on the draft.
This work
is supported by the European Union's Horizon 2020
research and innovation programme under the Marie
Sklodowska-Curie grant agreement No 895958.

\appendix

\section{Conventions}
\label{secConventions}

We work with the action $S_{\textrm{YM}}$ in the conventions of \cite{Erickson:1999qv,Erickson:2000af} and appendix A of \cite{Jiang:2019xdz}. It is useful to report the interactions (a)-(c) of section \ref{secPCGG} in the form suited for appendix \ref{secETdetails}.
\begingroup \allowdisplaybreaks 
\begin{flalign}
\label{eq34}
&
\left\langle \Phi_{I_{1}}^{A_{1}}\left(x_{1}\right)\Phi_{I_{2}}^{A_{2}}\left(x_{2}\right)\right\rangle _{\textrm{(a)}}
=-{g_{\textrm{YM}}^4 N}\delta_{I_{1}I_{2}}
\\
&~\times \left(\delta^{A_{1}A_{2}}-\delta^{A_{1}N^{2}}\delta^{A_{2}N^{2}}\right)
\left(Y_{x_{1}x_{1}x_{2}}+Y_{x_{1}x_{2}x_{2}}\right)
\nonumber
\\
\label{eq35}
&
\left\langle \Phi_{I_{1}}^{A_{1}}\left(x_{1}\right)\Phi_{I_{2}}^{A_{2}}\left(x_{2}\right)\Phi_{I_{3}}^{A_{3}}\left(x_{3}\right)\Phi_{I_{4}}^{A_{4}}\left(x_{4}\right)\right\rangle _{\textrm{(b)}}
\\
&
~={g_{\textrm{YM}}^6} \left[-\left(f^{AA_{1}A_{3}}f^{AA_{2}A_{4}}+f^{AA_{1}A_{4}}f^{AA_{2}A_{3}}\right)\delta_{I_{1}I_{2}}\delta_{I_{3}I_{4}}\right.
\nonumber
\\
&
~+\left(f^{AA_{1}A_{4}}f^{AA_{2}A_{3}}-f^{AA_{1}A_{2}}f^{AA_{3}A_{4}}\right)\delta_{I_{1}I_{3}}\delta_{I_{2}I_{4}}
\nonumber
\\
&
~\left.+\left(f^{AA_{1}A_{2}}f^{AA_{3}A_{4}}+f^{AA_{1}A_{3}}f^{AA_{2}A_{4}}\right)\delta_{I_{1}I_{4}}\delta_{I_{2}I_{3}}\right]X_{x_{1}x_{2}x_{3}x_{4}}
\nonumber
\\
\label{eq36}
&
\left\langle \Phi_{I_{1}}^{A_{1}}\left(x_{1}\right)\Phi_{I_{2}}^{A_{2}}\left(x_{2}\right)\Phi_{I_{3}}^{A_{3}}\left(x_{3}\right)\Phi_{I_{4}}^{A_{4}}\left(x_{4}\right)\right\rangle _{\textrm{(c)}}
\\
&=
{g_{\textrm{YM}}^6}\left[f^{AA_{1}A_{2}}f^{AA_{3}A_{4}}\delta_{I_{1}I_{2}}\delta_{I_{3}I_{4}}I_{x_{1}x_{2}}I_{x_{3}x_{4}}F_{x_{1}x_{2},x_{3}x_{4}}\right.
\nonumber
\\
&
~+f^{AA_{1}A_{3}}f^{AA_{2}A_{4}}\delta_{I_{1}I_{3}}\delta_{I_{2}I_{4}}I_{x_{1}x_{3}}I_{x_{2}x_{4}}F_{x_{1}x_{3},x_{2}x_{4}}
\nonumber
\\
&
~\left.+f^{AA_{1}A_{4}}f^{AA_{2}A_{3}}\delta_{I_{1}I_{4}}\delta_{I_{2}I_{3}}I_{x_{1}x_{4}}I_{x_{2}x_{3}}F_{x_{1}x_{4},x_{2}x_{3}}\right]
\nonumber
\end{flalign}
\endgroup

\section{Details on diagrammatic approach}
\label{secPCGGdetails}

The tree-level and one-loop parts of the partially-contracted giant gravitons operator are given by \cite{Jiang:2019xdz}
\begin{flalign}
\label{eq37}
&\mathcal{G}_{L}^{\left(0\right)} (x_1,x_2)
=
\frac{\left(N-L\right)!}{\left(L!\right)^{2}}\delta_{i_{1}\dots i_{L}}^{l_{1}\dots l_{L}}\delta_{k_{1}\dots k_{L}}^{j_{1}\dots j_{L}}
\\
&
~~\times \left(y_{i}\cdot\Phi\right)_{\:j_{1}}^{i_{1}}\dots\left(y_{i}\cdot\Phi\right)_{\:j_{L}}^{i_{L}}\left(y_{j}\cdot\Phi\right)_{\:l_{1}}^{k_{1}}\dots\left(y_{j}\cdot\Phi\right)_{\:l_{L}}^{k_{L}}\,,
\nonumber
\\
\label{eq38}
&\mathcal{G}_{L}^{\left(1\right)} (x_1,x_2)
=
g_{\textrm{YM}}^2 \frac{Y_{112}+Y_{122}}{I_{12}}
\left[
 L\left(L+1-N\right)
 \right.
\\
&
~~\times \mathcal{G}_{L}^{(0)} (x_1,x_2)
+\frac{\left(N-L\right)!}{\left(L!\right)^{2}}
M_{i_{1}\dots i_{L};k_{1}\dots k_{L}}^{l_{1}\dots l_{L};j_{1}\dots j_{L}}
\nonumber
\\
&
~~\left. \times \left(y_{i}\cdot\Phi\right)_{\:j_{1}}^{i_{1}}\dots\left(y_{i}\cdot\Phi\right)_{\:j_{L}}^{i_{L}}\left(y_{j}\cdot\Phi\right)_{\:l_{1}}^{k_{1}}\dots\left(y_{j}\cdot\Phi\right)_{\:l_{L}}^{k_{L}}\right]\,.
\nonumber
\end{flalign}
The relevant definitions are in main text and we define $M_{i_{1}\dots i_{L};k_{1}\dots k_{L}}^{l_{1}\dots l_{L};j_{1}\dots j_{L}}=\delta_{i_{1}\dots i_{L}i}^{l_{1}\dots l_{L}l}\delta_{k_{1}\dots k_{L}l}^{j_{1}\dots j_{L}i}-N\delta_{i_{1}\dots i_{L}}^{l_{1}\dots l_{L}}\delta_{k_{1}\dots k_{L}}^{j_{1}\dots j_{L}}$. Scalars carrying the index $i$ are in the position $x_1$ and the others  in $x_2$.

\paragraph{Tree level}
The Wick contractions produce
\begin{flalign}
&{G}_4
=
\frac{1}{\left(N!\right)^{4}}\sum_{\ell_{12}=0}^{N}\sum_{\ell_{13}=0}^{N-\ell_{12}}{N \choose \ell_{12}\:\ell_{13}\:\ell_{14}}{N \choose \ell_{12}\:\ell_{23}\:\ell_{24}}
\nonumber
\\
\label{eq7}
&
\times {N \choose \ell_{13}\:\ell_{23}\:\ell_{34}}{N \choose \ell_{14}\:\ell_{24}\:\ell_{34}}\prod_{i=1}^{4}\prod_{j>i}^{4}\ell_{ij}!\,(\Delta_{ij})^{\ell_{ij}}
\\
&
\times \left[\left(\ell_{12}!\right)^{4}\delta_{k_{1}\dots k_{\ell_{13}}m_{1}\dots m_{\ell_{14}}}^{r_{1}\dots r_{\ell_{13}}p_{1}\dots p_{\ell_{14}}}\delta_{q_{1}\dots q_{\ell_{13}}o_{1}\dots o_{\ell_{14}}}^{l_{1}\dots l_{\ell_{13}}n_{1}\dots n_{\ell_{14}}}\right.
\nonumber
\\
&
~~~~~ \left.\delta_{l_{1}\dots l_{\ell_{13}}p_{1}\dots p_{\ell_{14}}}^{q_{1}\dots q_{\ell_{13}}m_{1}\dots m_{\ell_{14}}}\delta_{r_{1}\dots r_{\ell_{13}}n_{1}\dots n_{\ell_{14}}}^{k_{1}\dots k_{\ell_{13}}o_{1}\dots o_{\ell_{14}}}\right]
+O\left(g_{\textrm{YM}}^{4N+2}\right)
\,.
\nonumber
\end{flalign}
Multinomial coefficients count the number of ways to split the $N$ scalars of a determinant into three sets (in preparation for the contractions), while factorials count the pairwise contractions of scalars that create the same bundle of $\ell_{ij}$ propagators. Contractions leave a product of $\epsilon$-symbols, here rewritten as generalized Kronecker deltas. This reduces to a purely counting problem because index permutations (e.g. $r_1\leftrightarrow r_2$) do not flip the sign of \eqref{eq7}. We view a sequence of homonymous indices as a set (e.g. ${\{r_i\}_{i=1,\dots, N}}$) and solve in two steps. First, the properties of $\delta$ fix the intersections of the sets (e.g. some $r$'s can equal some $k$'s, but not any $p$'s). This leads to sum over the number of elements in the intersection sets, whose traces are still visible in the final result \eqref{eq25}. Second, there are combinatorial factors related to these index identifications and the number of ways of assigning numerical values to all indices. 

\paragraph{One loop} The operator \eqref{eq38} encapsulates the one-loop diagrams between two determinants and can be used to quantify the two-determinant interactions
\begin{gather}
\sum_{\ell_{13}=1}^{N-1}
\Delta_{12}^{N-\ell_{12}} \Delta_{34}^{N-\ell_{34}}
\left\langle \mathcal{G}_{\ell_{13}}^{(1)}\left(x_{1},x_{2}\right)\mathcal{G}_{\ell_{13}}^{{(0)}}\left(x_{3},x_{4}\right)\right\rangle+\textrm{perm.}\,,
\end{gather}
where it is understood that Wick contractions are in free theory and cannot occur between $\Phi(x_1)$ and $\Phi(x_2)$ and between $\Phi(x_3)$ and $\Phi(x_4)$.  Three- and four-determinant interactions are obtained by performing free and one-loop contractions \eqref{eq34}-\eqref{eq36} among the determinants.

\section{Details on effective theory approach}
\label{secETdetails}
There is a shortcut that allows to recycle the knowledge of the basic interactions \eqref{eq34}-\eqref{eq36} and bypass the tedious construction of $\mathcal{O}_{a}^S$, $\mathcal{O}_{b}^S$ and $\mathcal{O}_{c}^S$ in step 2 in the main text. We exemplify it for the first one: the idea is to engineer the effective operator
\begin{flalign}
\label{eq39}
&\!\!\mathcal{O}_{a}
=
-\frac{\left(g_{\textrm{YM}}^{2}\right)^{-2}}{2!}g_{\textrm{YM}}^{4}N\int_{{y_{1},y_{2}}}2\left(Y_{y_{1}y_{1}y_{2}}+Y_{y_{1}y_{2}y_{2}}\right)
\\
&
\square_{y_{1}}\square_{y_{2}}\left[\textrm{tr}\left(\Phi^{I}\left(y_{1}\right)\Phi^{I}\left(y_{2}\right)\right)-\frac{\textrm{tr}\left(\Phi^{I}\left(y_{1}\right)\right)\textrm{tr}\left(\Phi^{I}\left(y_{2}\right)\right)}{N}\right]
\nonumber
\end{flalign}
such that
$\left\langle \Phi_{I_{1}}^{a_{1}}\left(x_{1}\right)\Phi_{I_{2}}^{a_{2}}\left(x_{2}\right)\mathcal{O}_{a}\right\rangle$ gives \eqref{eq34}. Indeed, contractions produce $2!$ propagators, which are removed using $\square_{x}I_{xy}=-\delta^{\left(4\right)}\left(x-y\right)$ and placing ${\left(g_{\textrm{YM}}^{2}\right)^{-2}}/{2!}$; self-contractions within $\mathcal{O}_{a}$ are understood to be discarded. Taking $\mathcal{O}_{a}$ as a starting point for path integration is very advantageous because this task becomes trivial as in free theory: one-loop interactions among scalars are already encoded into \eqref{eq39}, so one can straightforwardly replace $\Phi^I\to S^I$ in $\mathcal{O}_{a}$. The result coincides precisely with $\mathcal{O}_{a}^S$ given in the main text.

\bibliography{LetterRef}
\end{document}